\documentstyle[12pt]{ioplppt}      % use this for preprint style
\begin{document}
%\jl{6}
\title{Asymptotic expansion coefficients of the heat kernel in Riemann-Cartan space\footnote{This article is that the explicit form of $[a_q]\ (q=1,\cdots,5)$ is added to the proceeding of the 9th Marcel Grossmann meeting}}
\author{S. Yajima, Y. Higasida, K. Kawano and S.-I. Kubota}
\address{Department of Physics, Kumamoto University, \\
2-39-1 Kurokami, Kumamoto 860-8555, Japan\\
E-mail: yajima@aster.sci.kumamoto-u.ac.jp}
\begin{abstract}
By applying the covariant Taylor expansion method, the fifth lower coefficients the  asymptotic expansion of the heat kernel associated with a fermion of spin 1/2 in Riemann-Cartan space are manifestly given. These coefficients in Riemann-Cartan space is derived from those obtained in Riemannian space by simple replacements.
\end{abstract}
%
%\pacs{0240, 0450, 0462, 1110K}
%\maketitle
%
%\eqnobysec
%%%%%%%%%%%%%%%%%%%%%%%%%%%%%%%%%%%%%%%
\section{Introduction}%%% SECTION 1 %%%
%%%%%%%%%%%%%%%%%%%%%%%%%%%%%%%%%%%%%%%
Motivated by studying all one-loop quantities (such as the effective action, zeta function, Green functions, anomalies, etc.) in quantum field theory and supergravity, the heat kernel $K^{(d)}(x,x')$ in $d$ dimensions defined by equations
\begin{eqnarray}
&& {\partial \over \partial t} K^{(d)}(x,x';t) = - H K^{(d)}(x,x';t),   \nonumber \\
&& K^{(d)}(x,x';0)= {\bf 1} \vert h(x) \vert^{-1} \vert h(x') \vert^{-1} \delta^{(d)}(x,x'),  \label{eq:hk}
\end{eqnarray}
has been studied by many authors. Here $\delta^{(d)}(x,x')$ is the $d$ dimensional invariant $\delta$-function, 
and ${\bf 1}=\{\delta^A_{\ \ B}\}$, $h=\det{h^a_{\ \ \mu}}$, where $h^a_{\ \ \mu}$ is a vielbein. The most general elliptic second order differential operator $H$ for a fermion $\psi=\{\psi^A(x)\}$ of spin 1/2 in curved space with the torsion, i.e. in Riemann-Cartan space is expressed in the following form,
\begin{equation}
  H = D_{\mu} D^{\mu} + 2 Q^{\mu} D_{\mu} + Z  
    = \tilde{D}_{\mu} \tilde{D}^{\mu} + X  \label{eq:dirac}
\end{equation}
with
\begin{eqnarray}
&& \tilde{D}_{\mu} = D_{\mu} + Q_{\mu},\qquad  
D_{\mu} = \partial_{\mu} + {1 \over 4}(\omega_{ab\mu} 
+ {1 \over 2} h_a^{\ \alpha} h_b^{\ \beta} C_{\alpha\beta\mu}) \gamma^{ab} + iA_{\mu}, \nonumber \\
&&  Q_{\mu} = {1 \over 4} \gamma^{\alpha\beta} C_{\alpha\beta\mu}, \qquad
X = Z - Q^{\mu}_{\ :\mu} - Q_{\mu} Q^{\mu}, \qquad 
h^a_{\ \nu:\mu} \equiv \tilde{D}_{\mu} h^a_{\ \nu} = 0, \nonumber \\
&& \{ \gamma^a,\gamma^b \} = 2 \eta^{ab}, \quad \gamma^{ab} 
= {1 \over 2}[\gamma^a, \gamma^b], \quad \gamma^{\mu} 
= h_a^{\ \ \mu} \gamma^a,  
\end{eqnarray}
where $\omega_{ab\mu}$ is the Ricci coefficient of rotation, $C_{\alpha\beta\mu}$ a totally antisymmetric torsion tensor, and $A_{\mu}$ an arbitrary vector gauge field. The local quantity $Z$ stands for matrices for the spinor with non-differential operator. The metric tensor is given by $g_{\mu\nu} = h^a_{\ \mu} h^b_{\ \nu} \eta_{ab}$ 
with $\eta_{ab}={\rm diag}(-1,-1,\cdots,-1)$. 

The commutation relation of $\tilde{D}_{\mu}$ is given by
\begin{equation}
[\tilde{D}_{\mu},\tilde{D}_{\nu}] \psi = (\tilde{\Lambda}_{\mu\nu} 
+ C^{\rho}_{\ \ \mu\nu} \tilde{D}_{\rho}) \psi \label{eq:com}
\end{equation} 
with
\begin{eqnarray}
&& \tilde{\Lambda}_{\mu\nu} = \Lambda_{\mu\nu} + \nabla_{\mu}(C+Q)_{\nu} 
- \nabla_{\nu} (C+Q)_{\mu} + [(C+Q)_{\mu}, (C+Q)_{\nu}], \nonumber \\
&& \Lambda_{\mu\nu} = {1 \over 4} \gamma^{\alpha\beta} R_{\alpha\beta\mu\nu} 
+ i F_{\mu\nu},\qquad
C_{\mu} = {1 \over 8} \gamma^{\alpha\beta} C_{\alpha\beta\mu},
   \label{eq:curvature}
\end{eqnarray}
where $R_{\alpha\beta\mu\nu}$ and $F_{\mu\nu}$ are the curvature tensor in Riemannian space and the field strength of $A_{\mu}$ respectively, and $\nabla_{\mu}$ means the Riemannian covariant differentiation.

%%%%%%%%%%%%%%%%%%%%%%%%%%%%%%%%%%%%%%%
\section{Explicit form of asymptotic expansion coefficients}%%% SECTION 2 %%%
%%%%%%%%%%%%%%%%%%%%%%%%%%%%%%%%%%%%%%%
One can solve (\ref{eq:hk}) by the expression 
according to the De Witt's ansatz,
\begin{equation}
 K^{(d)}(x,x';t) \sim {\Delta^{1/2}(x,x') \over (4 \pi t)^{d/2}} 
{\rm exp}({\sigma(x,x') \over 2t}) \sum_{q=0}^{\infty} a_q(x,x') t^q,  
\end{equation}
where $\sigma(x,x')$ and $\Delta(x,x')$ are the geodesic distance and the Van Vleck-Morette determinant between $x$ and $x'$, respectively. The asymptotic expansion coefficients $[a_q]$ of the heat kernel are obtained as the coincidence limit ($x \rightarrow x'$) of $a_q(x,x')$. The explicit expressions of $[a_q]\ (q=1,\cdots,5)$ are written in the following form~\cite{rf:1},
\begin{eqnarray}
\fl {[a_1]} = P = - ({1 \over 6} R {\bf 1} + X),   \\
\fl {[a_2]} = -{1 \over 6} M_{(2)} + {1 \over 2} P^2 \nonumber \\
\fl \phantom{[a_2]} = {1 \over 12} \tilde{\Lambda}_{\mu\nu}
\tilde{\Lambda}^{\mu\nu}
+ {1 \over 180}(R_{\mu\nu\rho\sigma} R^{\mu\nu\rho\sigma}
-R_{\mu\nu} R^{\mu\nu}){\bf 1} \nonumber \\
\fl \phantom{[a_2] = } + {1 \over 6}({1 \over 5}R {\bf 1} + X)_{!\mu}^{\ \ \mu}
+ {1 \over 2}({1 \over 6}R {\bf 1} + X)^2, \\
\fl {[a_3]} = {1 \over 60} M_{(4)} - {1 \over 12}\{P,M_{(2)}\}
- {1 \over 12} (P_{!\alpha} -{1 \over 3} \tilde{J}_{\alpha})
(P^{!\alpha} +{1 \over 3} \tilde{J}^{\alpha}) + {1 \over 6}P^3  \nonumber \\
\fl \phantom{[a_3] = } -{1 \over 60} X_{!\mu\ \ \nu}^{\ \ \mu\ \ \nu}
- {1 \over 12} (XX_{!\mu}^{\ \ \mu} + X_{!\mu}^{\ \ \mu}X
+ X_{!\mu}X^{!\mu} ) -{1 \over 6}({1 \over 6}R {\bf 1} + X)^3 \nonumber \\
\fl \phantom{[a_3] = } - {1 \over 30} \{ X , \tilde{\Lambda}_{\mu\nu}
\tilde{\Lambda}^{\mu\nu} \}
- {1 \over 60} \tilde{\Lambda}_{\mu\nu} X \tilde{\Lambda}^{\mu\nu}
- {1 \over 60} [\tilde{J}^{\mu}, X_{!\mu}] \nonumber \\
\fl \phantom{[a_3] = } - {1 \over 36} R X_{!\mu}^{\ \ \mu} 
- {1 \over 30} R_{,\mu} X^{!\mu} 
- {1 \over 30} R_{,\mu}^{\ \ \mu} X 
- {1 \over 90} R^{\mu\nu} X_{!\mu\nu} \nonumber \\
\fl \phantom{[a_3] = } + {1 \over 180} (R_{\mu\nu} R^{\mu\nu}
- R_{\mu\nu\rho\sigma}R^{\mu\nu\rho\sigma})X \nonumber \\
\fl \phantom{[a_3] = } - {1 \over 10} \tilde{\Lambda}_{\mu}^{\ \ \nu}
\tilde{\Lambda}_{\nu}^{\ \ \rho} \tilde{\Lambda}_{\rho}^{\ \ \mu}
+ {1 \over 30} \{ \tilde{\Lambda}_{\mu\nu} , \tilde{J}^{\mu!\nu} \}
- {1 \over 45} \tilde{\Lambda}_{\mu\nu!\rho}
\tilde{\Lambda}^{\mu\nu!\rho} - {1 \over 180} \tilde{J}_{\mu}
\tilde{J}^{\mu} \nonumber \\
\fl \phantom{[a_3] = } - {1 \over 72} R \tilde{\Lambda}_{\mu\nu} \tilde{\Lambda}^{\mu\nu}
- {1 \over 18} R^{\mu\nu} \tilde{\Lambda}_{\mu\rho}
\tilde{\Lambda}_{\mu}^{\ \ \rho} + {1 \over 60} R^{\mu\nu\rho\sigma}
\tilde{\Lambda}_{\mu\nu} \tilde{\Lambda}_{\rho\sigma} \nonumber \\
\fl \phantom{[a_3]} + {\bf 1} \Big( 
- {1 \over 280} R_{,\mu\ \ \nu}^{\ \ \mu\ \ \nu}
- {1 \over 420} R^{\mu\nu} R_{,\mu\nu} + {1 \over 630} R^{\mu\nu}
R_{\mu\nu,\rho}^{\ \ \ \ \ \rho} - {17 \over 5040} R_{,\mu}
R^{,\mu} \nonumber \\
\fl \phantom{[a_3] = } - {1 \over 180} R R_{,\mu}^{\ \ \mu}
- {1 \over 105} R^{\mu\rho\nu\sigma} R_{\mu\nu,\rho\sigma}
+ {1 \over 2520} R_{\mu\nu,\rho} R^{\mu\nu,\rho}
+ {1 \over 1260} R_{\mu\nu,\rho} R^{\mu\rho,\nu} \nonumber \\
\fl \phantom{[a_3] = } - {1 \over 560} R_{\mu\nu\rho\sigma,\lambda}
R^{\mu\nu\rho\sigma,\lambda}
- {1 \over 5670} R^{\mu}_{\ \ \nu} R^{\nu}_{\ \ \rho}
R^{\rho}_{\ \ \mu} + {1 \over 1890} R_{\mu\nu} R_{\rho\sigma}
R^{\mu\rho\nu\sigma} \nonumber \\
\fl \phantom{[a_3] = } - {1 \over 270} R_{\mu\nu} R^{\mu\lambda\rho\sigma}
R^{\nu}_{\ \lambda\rho\sigma}
+ {4 \over 2835} R_{\mu\nu\rho\sigma} R^{\mu\nu\kappa\lambda}
R^{\rho\sigma}_{\ \ \ \kappa\lambda} \nonumber \\
\fl \phantom{[a_3] = } + {22 \over 2835} R_{\mu\rho\nu\sigma} R^{\mu\kappa\nu\lambda} R^{\rho\ \sigma}_{\ \kappa\ \lambda} 
+ {1 \over 1080} R
(R_{\mu\nu} R^{\mu\nu} - R_{\mu\nu\rho\sigma}R^{\mu\nu\rho\sigma}) \Big),
 \\
\fl {[a_4]} = -{1 \over 840} M_{(6)}
- {1 \over 40}\{P^2,M_{(2)}\}
- {1 \over 30} P M_{(2)} P + {1 \over 120}\{P,M_{(4)}\} \nonumber \\
\fl \phantom{[a_4] = } - {1 \over 30} \{ P, (P_{!\mu} -{1 \over 3} \tilde{J}_{\mu})
(P^{!\mu} +{1 \over 3} \tilde{J}^{\mu}) \} - {1 \over 60}
(P_{!\mu} -{1 \over 3} \tilde{J}_{\mu}) P 
(P^{!\mu} +{1 \over 3} \tilde{J}^{\mu}) \nonumber \\
\fl \phantom{[a_4] = } + {1 \over 60} (P_{!\mu} -{1 \over 3} \tilde{J}_{\mu})
(\tilde{\Lambda}^{\mu\nu} - {1 \over 3} R^{\mu\nu}{\bf 1})
(P_{!\nu} +{1 \over 3} \tilde{J}_{\nu})
+ {1 \over 72} M_{(2)}^2 \nonumber \\
\fl \phantom{[a_4] = } + {1 \over 60} \big((P_{!\mu} -{1 \over 3} \tilde{J}_{\mu})
(S^{\mu} +{1 \over 5} Y^{\mu})
+ (S_{\mu} -{1 \over 5} Y_{\mu})(P^{!\mu}
+{1 \over 3} \tilde{J}^{\mu})\big) \nonumber \\
\fl \phantom{[a_4] = } + {1 \over 90} 
(V_{\mu\nu} -{1 \over 2}\tilde{J}_{(\mu!\nu)})
(V^{\mu\nu} +{1 \over 2}\tilde{J}^{(\mu!\nu)}) 
+ {1 \over 24}P^4, \\
\fl {[a_5]} = {1 \over 120} P^5
-{1 \over 180}\{P^3,M_{(2)}\}
-{1 \over 120}P\{P,M_{(2)}\}P
+{1 \over 420}\{P^2,M_{(4)}\}          \nonumber \\
\fl \phantom{[a_5] = }  +{1 \over 280}PM_{(4)}P
-{1 \over 1680}\{P,M_{(6)}\}
+{13 \over 2520}\{P,M^2_{(2)}\}  \nonumber \\
\fl \phantom{[a_5] = } +{1 \over 280}M_{(2)}PM_{(2)}
-{1 \over 720}\{M_{(2)},M_{(4)}\}  \nonumber \\
\fl \phantom{[a_5] = } 
-{1 \over 120}\{P^2,(P_{!\mu}-{1 \over 3}\tilde{J}_{\mu})
(P^{!\mu}+{1 \over 3}\tilde{J}^{\mu})\}
-{1 \over 180}\{P,(P_{!\mu}-{1 \over 3}\tilde{J}_{\mu})P
(P^{!\mu}+{1 \over 3}\tilde{J}^{\mu})\}   \nonumber \\
\fl \phantom{[a_5] = } 
-{1 \over 90}P(P_{!\mu}-{1 \over 3}\tilde{J}_{\mu})
(P^{!\mu}+{1 \over 3}\tilde{J}^{\mu})P
-{1 \over 360}(P_{!\mu}-{1 \over 3}\tilde{J}_{\mu})
P^2(P^{!\mu}+{1 \over 3}\tilde{J}^{\mu})      \nonumber \\
\fl \phantom{[a_5] = } +{2 \over 315}
\big(P(P_{!\mu}-{1 \over 3}\tilde{J}_{\mu})
(S^{\mu}+{1 \over 5}Y^{\mu}) + (S^{\mu}-{1 \over 5}Y^{\mu})
(P_{!\mu}+{1 \over 3}\tilde{J}_{\mu})P\big)  \nonumber \\
\fl \phantom{[a_5] = } +{1 \over 140}\big((P_{!\mu}-{1 \over 3}\tilde{J}_{\mu})
(S^{\mu}+{1 \over 5}Y^{\mu})P + P(S^{\mu}-{1 \over 5}Y^{\mu})
(P_{!\mu}+{1 \over 3}\tilde{J}_{\mu})\big)    \nonumber \\
\fl \phantom{[a_5] = } +{1 \over 315}\big((S^{\mu}-{1 \over 5}Y^{\mu})P
(P_{!\mu}+{1 \over 3}\tilde{J}_{\mu})
+ (P_{!\mu}-{1 \over 3}\tilde{J}_{\mu})P
(S^{\mu}+{1 \over 5}Y^{\mu})\big)  \nonumber \\
\fl \phantom{[a_5] = } +{1 \over 210}
\{P,(V_{\mu\nu}-{1 \over 2}\tilde{J}_{(\mu!\nu)})
(V^{\mu\nu}+{1 \over 2}\tilde{J}^{(\mu!\nu)})\}   \nonumber \\
\fl \phantom{[a_5] = } +{1 \over 630}
(V_{\mu\nu}-{1 \over 2}\tilde{J}_{(\mu!\nu)})P
(V^{\mu\nu}+{1 \over 2}\tilde{J}^{(\mu!\nu)})    \nonumber \\
\fl \phantom{[a_5] = } +{13 \over 2520}\{M_{(2)},
(P_{!\mu}-{1 \over 3}\tilde{J}_{\mu})
(P^{!\mu}+{1 \over 3}\tilde{J}^{\mu})\} \nonumber \\
\fl \phantom{[a_5] = } +{1 \over 280}
(P_{!\mu}-{1 \over 3}\tilde{J}_{\mu})M_{(2)}
(P^{!\mu}+{1 \over 3}\tilde{J}^{\mu})  \nonumber \\
\fl \phantom{[a_5] = } +{1 \over 210}\big((P_{!\mu}-{1 \over 3}\tilde{J}_{\mu})
(P_{!\nu}-{1 \over 3}\tilde{J}_{\nu})
(V^{\mu\nu}+{1 \over 2}\tilde{J}^{(\mu!\nu)})  \nonumber \\
\fl \phantom{[a_5] = } \qquad+(V^{\mu\nu}-{1 \over 2}\tilde{J}^{(\mu!\nu)})
(P_{!\mu}+{1 \over 3}\tilde{J}_{\mu})
(P_{!\nu}+{1 \over 3}\tilde{J}_{\nu})\big) \nonumber \\
\fl \phantom{[a_5] = } +{1 \over 140}(P_{!\mu}-{1 \over 3}\tilde{J}_{\mu})
(V^{\mu\nu}+{1 \over 2}\tilde{J}^{(\mu!\nu)})(P_{!\nu}+{1 \over
3}\tilde{J}_{\nu})  \nonumber \\
\fl \phantom{[a_5] = } -{1 \over 360}(P_{!\mu}-{1 \over 3}\tilde{J}_{\mu})
(\tilde{\Lambda}^{\mu}_{\ \ \rho}-{1 \over 3}R^{\mu}_{\ \ \rho})
(\tilde{\Lambda}^{\rho}_{\ \ \nu}-{1 \over 3}R^{\rho}_{\ \ \nu})
(P^{!\nu}+{1 \over 3}\tilde{J}^{\nu})  \nonumber \\
\fl \phantom{[a_5] = } -{1 \over 315}\big((P_{!\mu}-{1 \over 3}\tilde{J}_{\mu})
(\tilde{\Lambda}^{\mu\nu}-{1 \over 3}R^{\mu\nu}{\bf 1})
(S_{\nu}+{1 \over 5}Y_{\nu})      \nonumber \\
\fl \phantom{[a_5] = } \qquad+(S_{\mu}-{1 \over 5}Y_{\mu})
(\tilde{\Lambda}^{\mu\nu}-{1 \over 3}R^{\mu\nu}{\bf 1})
(P_{!\nu}+{1 \over 3}\tilde{J}_{\nu})\big)  \nonumber \\
\fl \phantom{[a_5] = } -{1 \over 945}{\cal K}^{\mu\nu\rho\sigma}
(V_{\mu\nu}-{1 \over 2}\tilde{J}_{\mu!\nu})
(V_{\rho\sigma}+{1 \over 2}\tilde{J}_{\rho!\sigma}) \nonumber \\
\fl \phantom{[a_5] = } -{1 \over 315}
(V_{\mu\rho}-{1 \over 2}\tilde{J}_{(\mu!\rho)})
(\tilde{\Lambda}^{\mu}_{\ \ \nu}
-{1 \over 3}R^{\mu}_{\ \ \nu}{\bf 1})
(V^{\nu\rho}+{1 \over 2}\tilde{J}^{(\nu!\rho)})  \nonumber \\
\fl \phantom{[a_5] = } +{1 \over 180}\{P,(P_{!\mu}-{1 \over 3}\tilde{J}_{\mu})
(\tilde{\Lambda}^{\mu\nu}-{1 \over 3}R^{\mu\nu}{\bf 1})
(P_{!\nu}+{1 \over 3}\tilde{J}_{\nu})\} \nonumber \\
\fl \phantom{[a_5] = } +{1 \over 360}(P_{!\mu}-{1 \over 3}\tilde{J}_{\mu})
\{P,(\tilde{\Lambda}^{\mu\nu}-{1 \over 3}R^{\mu\nu}{\bf 1})\}
(P_{!\nu}+{1 \over 3}\tilde{J}_{\nu}) \nonumber \\
\fl \phantom{[a_5] = } 
+{1 \over 1260}\big((P_{!\mu}-{1 \over 3}\tilde{J}_{\mu})
(2R^{\mu(\rho,\sigma)}-R^{\rho\sigma,\mu}
-4\tilde{\Lambda}^{\mu(\rho!\sigma)})
(V_{\rho\sigma}+{1 \over 2}\tilde{J}_{(\rho!\sigma)})    \nonumber \\
\fl \phantom{[a_5] = } \qquad+(V_{\rho\sigma}
-{1 \over 2}\tilde{J}_{(\rho!\sigma)})
(2R^{\mu(\rho,\sigma)}-R^{\rho\sigma,\mu}
+4\tilde{\Lambda}^{\mu(\rho!\sigma)})
(P_{!\mu}+{1 \over 3}\tilde{J}_{\mu})\big)  \nonumber \\
\fl \phantom{[a_5] = } +{1 \over 280}
(P_{!\mu}-{1 \over 3}\tilde{J}_{\mu})e^{\mu\nu}
(P_{!\nu}+{1 \over 3}\tilde{J}_{\nu})  \nonumber \\
\fl \phantom{[a_5] = } 
-{1 \over 560}\big((P_{!\mu}-{1 \over 3}\tilde{J}_{\mu})
({\cal S}^{\mu}+{1 \over 7}{\cal Y}^{\mu})
+({\cal S}^{\mu}-{1 \over 7}{\cal Y}^{\mu})
(P_{!\mu}+{1 \over 3}\tilde{J}_{\mu})\big)    \nonumber \\
\fl \phantom{[a_5] = } -{1 \over 420}
\big((V^{\mu\nu}-{1 \over2}\tilde{J}^{(\mu!\nu)})
(U_{\mu\nu}+W_{\mu\nu})+(U_{\mu\nu}-W_{\mu\nu})
(V^{\mu\nu}+{1 \over2}\tilde{J}^{(\mu!\nu)})\big)    \nonumber \\
\fl \phantom{[a_5] = } -{17 \over 5040}
(S_{\mu}-{1 \over 5}Y_{\mu})(S^{\mu}+{1 \over 5}Y^{\mu})
-{1 \over 840}(S_{\mu\nu\rho}-{1 \over 5}Y_{\mu\nu\rho})
(S^{\mu\nu\rho}+{1 \over 5}Y^{\mu\nu\rho})    \nonumber \\
\fl \phantom{[a_5] = } +{1 \over 15120} M_{(8)},  \\ \nonumber
\end{eqnarray}
where

Here
\begin{eqnarray}
  &{\cal K}&_{\alpha\beta\mu_1\cdots\mu_n} \equiv
R_{\alpha(\mu_1\vert\beta\vert\mu_2,\mu_3\cdots\mu_n)}, \qquad
 {\cal K}^{\alpha\beta}_{\ \ \ \mu_1\cdots\mu_n(2k)}
\equiv {\cal K}^{\alpha\beta\quad\quad\quad\quad\ \
\lambda_1\cdots\lambda_k}_{\ \ \ \mu_1\cdots\mu_n\lambda_1
\cdots\lambda_k},  \nonumber \\
  &{\cal M}&_{\mu_1\cdots\mu_n} \equiv
{\cal K}^{\rho}_{\ \rho\mu_1\cdots\mu_n}
= R_{(\mu_1\mu_2,\mu_3\cdots\mu_n)},  \qquad
 {\cal M}_{\mu_1\cdots\mu_n(2k)} \equiv
{\cal M}_{\mu_1\cdots\mu_n\lambda_1\cdots\lambda_k}
^{\quad\quad\quad\quad\ \ \lambda_1\cdots\lambda_k},   \nonumber \\
 &\tilde{\Lambda}&^{\alpha}_{\ \ \mu_1\cdots\mu_n}
\equiv \tilde{\Lambda}^{\alpha}_{\ \ (\mu_1!\mu_2\cdots\mu_n)}, \qquad
\tilde{\Lambda}^{\alpha}_{\ \ \mu_1\cdots\mu_n(2k)}
\equiv \tilde{\Lambda}^{\alpha\quad\quad\quad\quad\ \ \ \lambda_1
\cdots\lambda_k}_{\ \ \mu_1\cdots\mu_n\lambda_1\cdots\lambda_k},  \nonumber
\\
 &X&_{\mu_1\cdots\mu_n} \equiv X_{!(\mu_1\cdots\mu_n)}, 
\end{eqnarray}
and the exclamation mark '$!\mu$' means the differentiation
of '$:\mu$' or $\tilde{D}_{\mu}$ in the case
that the torsion tensor in $\tilde{\Gamma}^{\lambda}_{\ \ \nu\mu}$
is eliminated,
\begin{eqnarray}
 I_{!\mu} &=& (\nabla + {\cal C} + Q)_{\mu} I =  I_{,\mu}
+ ({\cal C} + Q)_{\mu} I, \nonumber \\
 I_{!\mu\nu} &=& (\nabla + {\cal C} + Q)_{\nu} I_{!\mu}
=  I_{!\mu,\nu} + ({\cal C} + Q)_{\nu} I_{!\mu}, \nonumber \\
 \tilde{\Lambda}_{\mu\nu!\rho} &=& [(\nabla + {\cal C} + Q)_{\rho} ,
\tilde{\Lambda}_{\mu\nu}] =  \tilde{\Lambda}_{\mu\nu,\rho}
+ [({\cal C} + Q)_{\rho} , \tilde{\Lambda}_{\mu\nu}].
\end{eqnarray}
%

%%%%%%%%%%%%%%%%%%%%%%%%%%%%%%%%%%%%%%%
\section{Discussion}%%% SECTION 3 %%%
%%%%%%%%%%%%%%%%%%%%%%%%%%%%%%%%%%%%%%%
The covariant Taylor expansion~\cite{rf:2} for a function $f(x')$ at $x^{\mu}$ with respect to $\tilde{D}^{\mu'}\sigma(x,x')$ includes the expansion~\cite{rf:3} for the function with respect to the geodesic normal coordinate $y^{\mu}=x'^{\mu}- x^{\mu}$ satisfying 
\begin{eqnarray}
&& \tilde{\Gamma}^{\mu}_{\ \alpha\beta}(y) y^{\alpha} y^{\beta} 
= \Gamma^{\mu}_{\ \alpha\beta}(y) y^{\alpha} y^{\beta} =0,  \nonumber \\
&& y^{\mu} \tilde{\Omega}_{\mu}(x') \equiv y^{\mu} \{{1 \over 4} \gamma^{ab} 
\omega_{ab\mu} + C_{\mu} + Q_{\mu} + iA_{\mu} \}(y) =0.  
\label{eq:connection}
\end{eqnarray}
In the case that $x$ is close to $x'$, the conditions (\ref{eq:connection}) are obtained from $\tilde{D} \sigma^{:\mu}= \sigma^{:\mu}$ and $\sigma^{:\mu} \tilde{D}_{\mu} I = 0$, respectively, because $\sigma^{:\mu'} \approx y^{\mu}$, $I \approx {\bf 1}$ and $I_{:\mu} \approx 0$, where $I \equiv a_0$. The geodesic normal coordinate expansion for $h^a_{\ \ \mu}$, $h_a^{\ \ \mu}$, $g^{\mu\nu}$, $\log{\vert g \vert}$ ($g = \det{g_{\mu\nu}}$) and $\tilde{\Omega}_{\mu}$ at $x^{\mu}$ with respect to $y^{\mu}$ is expressed by the covariant expansion coefficients for $\tilde{\theta}^{\mu'}_{\ \ \nu} \equiv \sigma^{:\mu'}_{\ \ \ \nu}$, 
$\tilde{\beta}^{\mu}_{\ \ \nu'} \equiv (\tilde{\theta}^{-1})^{\mu}_{\ \ \nu'}$, 
$G^{\mu'\nu'} \equiv \tilde{\theta}^{\mu'}_{\ \ \alpha}
\tilde{\theta}^{\mu'\alpha}$, 
$\zeta \equiv \ln\{\det{(-\tilde{\beta}^{\mu'}_{\ \ \nu'})}\}^{-1/2}$ and 
${\cal A}_{\mu'} \equiv \tilde{\beta}^{\nu}_{\ \ \mu'} I^{-1} \tilde{D}_{\nu} I$ 
respectively~\cite{rf:2},
\begin{eqnarray}
&& h^a_{\ \ \mu}(x') = \sum^{\infty}_{n = 0} {1 \over n!} 
\tilde{\beta}^a_{\ \ \mu\mu_1 \cdots \mu_n}(x) y^{\mu_1} \cdots y^{\mu_n}, \nonumber \\
&& h_a^{\ \ \mu}(x') = \sum^{\infty}_{n = 0} {1 \over n!} 
\tilde{\theta}^{\mu}_{\ \ a\mu_1 \cdots \mu_n}(x) 
y^{\mu_1} \cdots y^{\mu_n}, \nonumber \\
&& g^{\mu\nu}(x') = \sum^{\infty}_{n = 0} {1 \over n!} 
G^{\mu\nu}_{\ \ \ \mu_1 \cdots \mu_n}(x) y^{\mu_1} \cdots y^{\mu_n}, \nonumber \\
&&  \log{\vert g \vert}(x') =  4 \sum^{\infty}_{n = 2} {1 \over n!}
\zeta_{\mu_1 \cdots \mu_n}(x) y^{\mu_1} \cdots y^{\mu_n}, \nonumber \\
&&  \tilde{\Omega}_{\mu}(x') = \sum^{\infty}_{n = 1} {1 \over n!}
\tilde{\cal A}_{\mu \mu_1 \cdots \mu_n} (x) y^{\mu_1} \cdots y^{\mu_n}.
\label{eq:expansion}
\end{eqnarray}
The expansion for $g^{\mu\nu}(x')$ and $\log{\vert g \vert}(x')$ are independent of the torsion tensor, though $h^a_{\ \ \mu}(x')$ and $h_a^{\ \ \mu}(x')$ depend on the torsion tensor. 
The torsion tensor is included in the expansion coefficients 
of $\tilde{\Omega}_{\mu}(x')$, in the form of $(C+Q)_{\mu}$, 
and appears in $\tilde{\Lambda}_{\mu\nu}$, $X$ 
and their derivatives in the coefficients $[a_q]$. 
In conclusion, the coefficients $[a_q]$ obtained by starting 
from Eq.~(\ref{eq:dirac}) in Riemann-Cartan space 
is derived from the final form of $[a_q]$ obtained 
by starting from $\nabla_{\mu} \nabla^{\mu} + Z$ in Riemannian space 
by the simple replacements~\cite{rf:4},
\begin{equation}
 Z \rightarrow X, \qquad 
\Lambda_{\mu\nu} \rightarrow \tilde{\Lambda}_{\mu\nu},\qquad 
\nabla_{\mu} \rightarrow \nabla_{\mu} + (C + Q)_{\mu}. 
\end{equation} 
%
%%%%%%%%%%%%%%%%%%%%%%%%%%%%%%%%%%%%%%%%%%%%%%%%%%%%%%%%%%%%%%%%%%%%%%%%%%%%%

%%%%%%%%%%%%%%%%%%%%%%%%%%%%%%%%%%%%%%%%%%
\section*{References}%%% REFERENCES %%%
%%%%%%%%%%%%%%%%%%%%%%%%%%%%%%%%%%%%%%%%%%

%%%%%%%%%%%%%%%%%%%%%%%%%%%%%%%%%%%%%%
\end{document}